\documentclass[twocolumn,showpacs,prb,floatfix]{revtex4}

\usepackage{graphicx}
\usepackage{dcolumn}
\usepackage{amsmath}
\usepackage{amssymb}

\def\beq{\begin{equation}}
\def\eeq{\end{equation}}
\def\nn{\nonumber}
\def\om{\omega}

\def\b{\beta}

\def\s{\sigma}

\def\d{\delta}
\def\dv{\delta_{4\pi}}

\def\rd{{\rm{d}}}
\def\vp{\varphi}
\def\vp{\mathbf{p}}

\def\vn{\mathbf{n}}
\def\vm{\mathbf{m}}
\def\vQ{\mathbf{Q}}

\def\vJ{\mathbf{J}}

\begin{document}
\title{Conductivity of a quasiperiodic system in two and three dimensions}

\author{ D.~Braak\footnote{email: Daniel.Braak@physik.uni-augsburg.de}
}
\affiliation{Theoretical Physics II, Institute for Physics,
University of Augsburg, 86135 Augsburg, Germany}

\date{\today}

\begin{abstract}
A generalization of the  
Aubry-Andr\'e model in two and
three dimensions is introduced which 
allows for quasiperiodic hopping terms
in addition 
to the  quasiperiodic site potentials. This
corresponds to  an array of interstitial impurities
within the periodic host crystal. The
resulting model is exactly solvable
and I compute the density of states and
the ac-conductivity. There is no mobility edge as in 
completely
disordered systems but the regular ac-conductivity and the
strongly reduced Drude weight indicate a precursor of the
Anderson transition as the Fermi energy goes from the 
center to the band edges.
\end{abstract}


\maketitle

\begin{figure}[h]
\begin{center}
\includegraphics[width=60mm,clip]{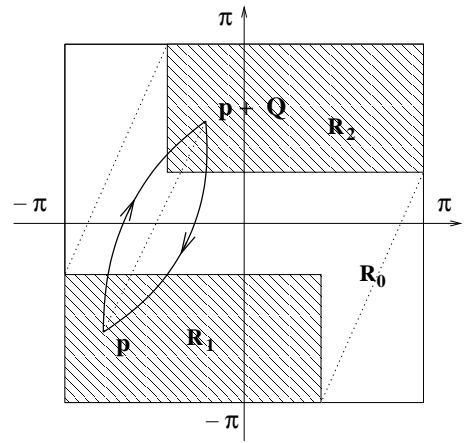}
\end{center}
\caption{The three regions $R_0,R_1,R_2$ for $d=2$.} \label{fig1}
\end{figure}

\begin{figure}[h]
\begin{center}
\includegraphics[width=80mm,clip]{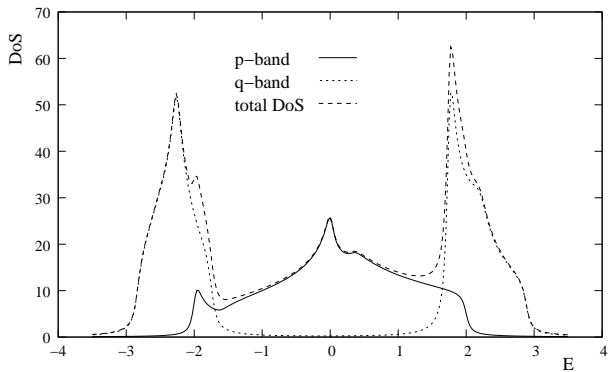}
\end{center}
\caption{Density of states for $d=2$. The potential is quasiperiodic
with
$Q_x=2\pi-4$ and $Q_y=2\pi-2\sqrt{2}$. The coupling constant is
  $g=2$. } \label{dos2d}
\end{figure}

\begin{figure}[h]
\begin{center}
\includegraphics[width=80mm,clip]{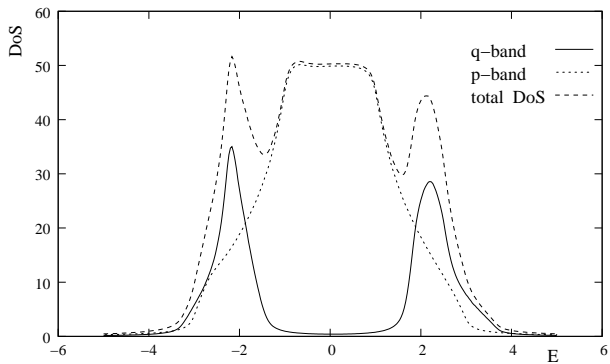}
\end{center}
\caption{DoS for $d=3$. Here, $Q_x=2\pi-4,Q_y=2\pi-2\sqrt{3}$ and
$Q_z=2\pi-\sqrt{5}$ with coupling  $g=2$.} \label{dos3d}
\end{figure}

\begin{figure}[h]
\begin{center}
\includegraphics[width=75mm,clip]{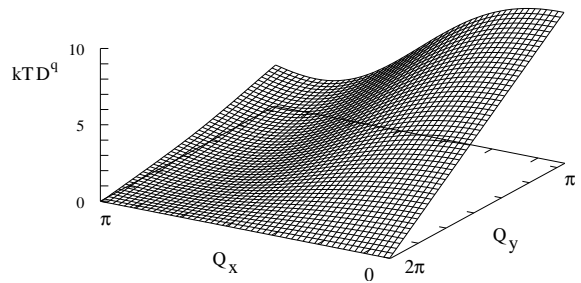}
\end{center}
\caption{Drude weight of the $q-$band $D^q_x/\b$ for $d=2$. The maximum at
  $Q_y=\pi$ and $Q_x=0$ corresponds to the Drude weight of the pure
  system $D_0$. The minimum along $Q_y=2\pi$ is due to the vanishing
  area of $R_1$ in this case.} \label{drude2d}
\end{figure}

\begin{figure}[h]
\begin{center}
\includegraphics[width=80mm,clip]{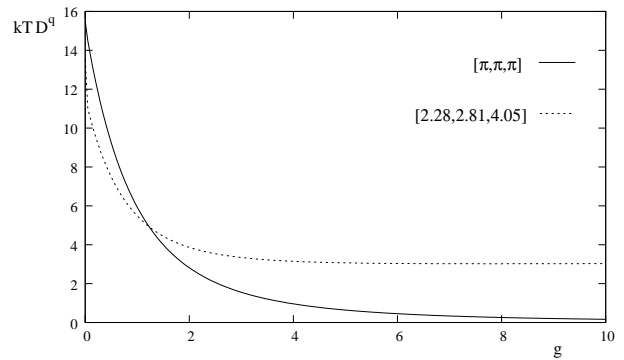}
\end{center}
\caption{Drude weight $D^q_x/\b$ for $d=3$ and two different vectors
  $\vQ$. The large coupling limit is zero for periodic $\vQ=(\pi,\pi,\pi)$ and
  nonzero
for quasiperiodic $\vQ=(2\pi-4,2\pi-2\sqrt{2},2\pi-\sqrt{5})$.} \label{drude3d}
\end{figure}

\begin{figure}[h]
\begin{center}
\includegraphics[width=80mm,clip]{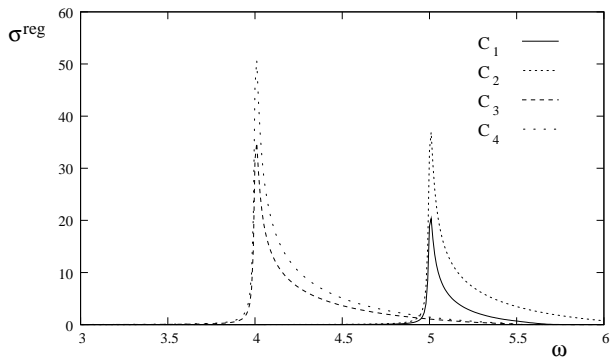}
\end{center}
\caption{$\s^{reg}_{xx}(\om)$ for various parameters: $C_3$ corresponds to
  $g=2$,$\b=1$,$\vQ=(2\pi-4,2\pi-3)$, $C_4$ has the same $g$ and
  $\vQ$ but $\b=5$, $C_2$ and $C_1$ both have $g=2.5$ and $\b=1$ but
  different $\vQ$: $(2\pi-4,2\pi-3)$ and
  $(2\pi-4.8,2\pi-2.5)$. Although $R_1$ has the same area in both
  cases, the ac-conductivity is quite different. $E_F=-2$ in all
  cases; the Fermi energy lies within the lower $q-$band.}    \label{rsigma}
\end{figure}

The description of electron motion in a non-periodic potential
within the single-particle approximation 
is an elementary problem in solid state
physics and still unsolved for the case of a macroscopic number of
impurities which break discrete translational invariance. 
Especially the phenomenon of Anderson localization in a tight-binding
model with uncorrelated random site potentials defies exact analytical
treatment up to now.
 Therefore, several attempts have been undertaken
to study models with  quasiperiodic potential, which constitute a
case intermediate between the perfect crystal and a fully disordered
system. 
The simplest realization of this situation is a 
tight-binding hamiltonian $H_1$ where the site energies vary
in a quasiperiodic fashion:
\beq
H_1 = H_0 +\sum_{\vm,\vn} g_{\vm}\d_{\vn\vm}|\vn\rangle\langle\vn|.
\label{tight}
\eeq
The $|\vn\rangle$ are local orbitals at site $\vn$ of a
$d-$dimensional hypercubic lattice, $H_0$ describes
the hamiltonian of the periodic lattice and $g_{\vm}$ is
a quasiperiodic function of $\vm$.\\
A possible choice for $g_{\vm}$ is
\beq
g_{\vm}= 2g\cos(\vQ\cdot\vm),
\label{AA}
\eeq
with a vector $\vQ$ whose components $Q_j$ are incommensurable with
$\pi$. This is the $d-$dimensional Aubry-Andr{\'e} model and
 unsolvable for $d>1$ despite its simplicity. Even in $d=1$
only very limited results can be obtained exactly, e.g. the
localization property of the eigenstates if $g >1$ \cite{aa}.\\
A  general criticism applies to all models of type (\ref{tight}):
The doping of the periodic host lattice with impurities is usually not of the
substitutional type but the additional atoms sit at interstitial
positions.
They modify not only the site
energies but change locally the orbitals and in turn the overlap
integrals
defining the hopping matrix elements. Furthermore, an empty impurity
orbital may serve as intermediate state for a hopping process
connecting distant sites of the host lattice. 
These effects may be phenomenologically taken into account by the following
generalization of (\ref{tight}):
\beq
H = H_1 + \sum_{\vm}\left(
\sum_{\vn,\vn'} g'_{\vm}(\vn-\vm,\vn'-\vm)|\vn\rangle\langle\vn'|
\right).
\label{extension} 
\eeq
The function $g'_{\vm}(\vn-\vm,\vn'-\vm)$   depends on the
position of the impurity $\vm$ and decays with growing distance
of the sites $\vn, \vn'$ from $\vm$. This generalization of 
(\ref{tight}) seems to be more complicated than the Aubry-Andr\'e model if 
both the $g_{\vm}$ and the $g'_{\vm}$ vary quasiperiodically with
$\vm$. It is therefore surprising that one may construct such
a model which is
completely analytically solvable and
has a solution in arbitrary dimensions.
``Complete solution'' means here that the hamiltonian is
diagonalizable and computation of the transport properties 
can be done analytically.
As in most non-trivial soluble models, 
the solvability rests here on a relation between the
quasiperiodic functions $g_{\vm}$ and   $g'_{\vm}$.  
The starting point is the
periodic hamiltonian
\beq
H_0 = -t\sum_{(\vn,\vn')} |\vn\rangle\langle\vn'| + {\rm h.c.}
\eeq
where $(\vn,\vn')$ denotes a pair of next neighbors on the
$d-$dimensional hypercubic lattice $\Lambda=\mathbb{Z}^d$ 
with lattice constant $a=1$.
Its matrix elements read in momentum space
\beq
H_0(\vp,\vp') = -2t\sum_{j=1}^d \cos(p_j)\d(\vp-\vp').
\eeq  
with $\vp,\vp'\in [-\pi,\pi]^d$.\\
We introduce two types of potentials related to the site
$\vm$:\\
I.\ A local potential term with matrix element $V_\vm(\vn,\vn')$
in position space:
\beq
V_\vm(\vn,\vn') = g_\vm(2\pi)^d\prod_{j=1}^d\d(n_j-n_j')\d(n_j-m_j).
\label{local}
\eeq
II.\ A non-local term:
\begin{eqnarray}
V'_\vm(\vn,\vn') = \nonumber\\ 
g'_\vm\left(\frac{2}{\pi}\right)^d
\prod_{j=1}^d\frac{(-1)^{n_j+n_j'}}{(n_j-m_j-1/2)(n_j'-m_j-1/2)}.
\label{nlocal}
\end{eqnarray}
The functional form of (\ref{nlocal}) agrees with the general
expression in (\ref{extension}) and can be interpreted as 
perturbative description of a hopping process, where the electron 
goes from site $\vn$ first to 
the impurity site $\vm + \mathbf{1/2}$ on the dual
lattice
and from there to the site $\vn'$. The technical reason for
this choice of the non-local term is the fact that now
both potentials have the same form in momentum space:
\begin{eqnarray}
V_\vm(\vp,\vp') &=& g_\vm\exp\left(i\vm\cdot(\vp'-\vp)\right)\\
V'_\vm(\vp,\vp') &=& g'_\vm\exp\left(i\widetilde{\vm}\cdot(\vp'-\vp)\right).
\label{Vmomentum}
\end{eqnarray}
with $\widetilde{m}_j=m_j+1/2$. Because both forms of the potential
factorize with respect to the dimension index $j$, also a mixed
type is possible, which is for some $j$ of type I and for the rest 
of type II. All of these possibilities can be parameterized through a
vector $\vm$ with integer and/or half-integer components, i.e.
$\vm\in \Lambda'=(\mathbb{Z}/2)^d$ and we may drop the primes
in (\ref{Vmomentum}). The generalization of the 
Aubry-Andr\'e model in $d$ dimensions reads then
\beq
H = H_0 + \frac{1}{(2\pi)^d}\sum_{\vm\in \Lambda'}V_\vm
\label{ham}
\eeq
and 
\beq
g_\vm = \left\{
\begin{array}{ll}
2g_{\rm loc}\cos(\vQ\cdot\vm), & \vm \in \Lambda \\
2g_{\rm nloc}\cos(\vQ\cdot\vm), & \vm \in \Lambda' \setminus \Lambda\  . 
\end{array}
\right. 
\label{q-per} 
\eeq
Here, $\vQ$ denotes some vector in $\mathbb{R}^d$ such that $Q_j/\pi$
are irrational. We may further assume that the numbers $Q_j/\pi$
are linearly independent over $\mathbb{Z}$. In this case, no point of
the lattice is equivalent to another. The original Aubry-Andr\'e model
is recovered for $g_{\rm nloc}=0$. We set in the following 
$g_{\rm loc}=g_{\rm nloc}=g$.
The potential term $V_\vQ=(2\pi)^{-d}\sum_{\vm\in\Lambda'} V_\vm$ can be rewritten
in momentum space:
\beq
V_\vQ(\vp,\vp') 
= g\left(\d_{4\pi}(\vp'-\vp+\vQ)+\d_{4\pi}(\vp'-\vp-\vQ)\right). \label{potmom}
\eeq
Here, $\dv(x)$ denotes the  delta-function with fundamental period $4\pi$: $\dv(x +
4n\pi) = \dv(x)$ for $n\in\mathbb{Z}$. The quasiperiodic potential
translates in momentum space to displaced delta-functions, which
however do not have  $2\pi-$periodicity but $4\pi-$periodicity due to the addition  
of the non-local terms of type II.
As a result, only finitely many points in momentum space are connected
through the potential term, even if $\vQ/\pi$ is irrational.\\ 
The simplest case is realized if just  two sites in momentum
space are connected.
This happens if one of the $Q_j$ lies in the interval
$[\pi,2\pi]$. The other components of $\vQ$ can be chosen at will.
The Brillouin zone splits into three regions $R_0,R_1$ and $R_2$,
where $R_0$ contains points, which are not affected by the potential (\ref{potmom})
at all, whereas $R_1$ and $R_2$ are mutually connected through the
potential:
\begin{eqnarray}
R_1 &=& \prod_{j=1}^d[-\pi,\pi-Q_j] \nn\\
R_2 &=& \prod_{j=1}^d[-\pi+Q_j,\pi] \label{regions}\\
R_0 &=& \prod_{j=1}^d[-\pi,\pi] \setminus (R_1 \cup R_2) \nn
\end{eqnarray} 
 (see Fig.\ref{fig1} for the case $d=2$).\\
It has to be emphasized that this splitting is equivalent to a
reduction of the Brillouin zone and concomitant generation of sub-bands
only if $\vQ/\pi$ is a rational vector. In the general case the model
does
not have a band structure.\\
Nevertheless, vectors $\vp$ from $R_1$ can be used to label the
two-dimensional invariant subspaces ${\cal{H}}_{\vp}$, which are
spanned by $|\vp\rangle$ and $|\vp+\vQ\rangle$.     
On these subspaces, the hamiltonian can be easily
diagonalized:
\beq
H\Big|_{{\cal{H}}_{\vp}}=\left(
\begin{array}{cc}
-2t\sum_j\cos(p_j) & g\\
g & -2t\sum_j\cos(p_j+Q_j)
\end{array} 
\right). 
\label{hamp}
\eeq
In the following we have set for simplicity $t=1/2$.
 Fig.~\ref{dos2d} and \ref{dos3d} give the density of
states in two and three dimensions, respectively. The total DoS is
composed from two ``bands'', where the $p-$band stems from the region
$R_0$, which is unaffected by the quasiperiodic potential and the eigenstates of
(\ref{ham})
are the pure momentum eigenstates $|\vp\rangle$ for $\vp\in R_0$. These are located in a region
around $E=0$, the band center. The $q-$band is gapped and contains the
eigenstates
$|\psi^\pm_\vp\rangle$ in ${\cal{H}}_{\vp}$ for each $\vp\in R_1$ and
energy eigenvalues $E^\pm_\vp$. It is located predominantly at the
band edges. \\
To compute the conductivity, we first note that
the $p-$band is not affected by the potential: The current commutes
with the hamiltonian and we get a diagonal ac-conductivity ($\b=(kT)^{-1}$): 
\beq
{\textrm{Re}}\left(\s_{kk}(\om)\right) = \pi D_0(\b)\d(\om) + \s^{reg}_{kk}(\om)
\eeq
with vanishing regular part, $\s^{reg}_{kk}(\om)=0$ and the Drude weight
$D_0(\b)$ of
the
pure system, which is only reduced because $R_0$ does not cover the
whole Brillouin zone.\\
In the $q-$band, however, we obtain a non-vanishing $\s^{reg}_{kk}$
because the current $\vJ$ does not commute with $H$ on the spaces  ${\cal{H}}_{\vp}$:
\beq
J_k\Big|_{{\cal{H}}_{\vp}}=e\left(
\begin{array}{cc}
\sin(p_k) & 0\\
0 & \sin(p_k+Q_k)
\end{array} 
\right). 
\label{current}
\eeq
The (anisotropic) Drude weight $D^q_k(\b)$ of the $q-$band is computed as
\beq
D^q_k(\b)=
\b\int_{R_1}\rd\vp\ 
n_F(E^\pm_\vp)|\langle\psi_\vp^\pm|J_k|\psi_\vp^\pm\rangle|^2
\label{Drude}
\eeq
with the Fermi-function $n_F(E)=(1+\exp\b(E-E_F))^{-1}$. $D^q_k$
approaches $D_0$ for  $Q_k\rightarrow 0,2\pi$,
i.e. when the potential becomes periodic in
$k-$direction. Fig.\ref{drude2d}
shows $D^q_x/\b$ for small temperatures and $E_F$ located in the gap
of the $q-$band as function of $Q_x$ and $Q_y$.
The Drude weight  $D^q_k$ approaches a nonzero large coupling limit
$g\rightarrow \infty$ with the
exception of the (rational) value $Q_k/\pi=1$, where the expectation
value of $J_k$ in the eigenstates of $H$ goes to zero.
This behavior is seen in Fig.\ref{drude3d}. 
Apart from $D^q_k$, which is reduced in the $q-$band due to scattering
from the quasiperiodic potential, the regular part of the
ac-conductivity does not vanish:
\begin{eqnarray}
\s^{reg}_{kk}(\om)=  
\pi\frac{1-e^{-\b\om}}{\om}\times \nonumber\\
\int_{R_1}\rd\vp\ n_F(E^-_\vp) 
|\langle\psi_\vp^-|J_k|\psi_\vp^+\rangle|^2\d(\om+E^-_\vp-E^+_\vp) 
\label{sigmareg}
\end{eqnarray}
with
$E_F\le 0$. 
Because $E^+_\vp-E^-_\vp$ is bounded from below by $2g$, the
ac-conductivity vanishes for frequencies below this threshold. 
$\s^{reg}_{xx}(\om)$ is plotted in Fig.\ref{rsigma} for various values
of $\beta$, $g$ and  $\vQ$ in $d=2$. The form of
$\s^{reg}(\om)$ in three dimensions is very similar.
\par
\noindent
We have introduced and solved a tight-binding model with quasiperiodic
local 
and non-local potentials 
in arbitrary dimension. The eigenstates  fall into two groups,
 where one group (states in the $p-$band around $E=0$) exhibits the properties of the pure system
but states at the band edges
($q-$band), although still extended, have different transport
 properties 
due to the influence of the quasiperiodic potential: a strongly
reduced Drude weight and non-zero regular ac-conductivity. This can be
interpreted as  a precursor of the Anderson transition expected for 
lattices with uncorrelated site potentials.
For irrational values of $Q_j/\pi$ the system has no band structure in
the usual sense.
Nevertheless, the density of states is absolute continuous and therefore
``band-like'' \cite{yuan}. This feature has been found
numerically in the two-dimensional labyrinth-tiling \cite{yuan}, the analogous
three-dimensional model \cite{cerov} and 
the square Fibonacci tiling \cite{mandel} for a certain range of parameters. 
The ``band-like'' spectra are
{\it prima facie} closer to the experimentally
observed quasiperiodic systems \cite{exp} than the models which
exhibit a singular continuous spectrum like most of the
one-dimensional examples \cite{theor}.
But this does not mean that the present model is  typical for
real quasicrystals.
The main objection to the physical relevance of the form
(\ref{nlocal}) for the non-local term is the following: the hopping  matrix
elements decay algebraically 
with distance from the impurity site
whereas the decay should be exponential. In this respect the model 
resembles the Lloyd model, which describes uncorrelated disorder with
a broad (lorentzian) distribution of the site potentials\cite{loy}.
However, only the disorder-averaged
 DoS can be analytically calculated in the Lloyd
model, not the transport properties. It is therefore only partially
solvable. Of course, the ``long-range'' nature 
of the quasiperiodic hopping term has a delocalizing effect on the
eigenstates and is probably the reason for the absence of a
mobility edge in the generalized Aubry-Andr\'e model.\\
Besides the conductivity, one may study the temporal behavior
of wave-packets initially located at a single site and
the associated (anomalous) diffusion of the electrons.  
This has been done numerically for several quasiperiodic
systems in two and three dimensions \cite{yuan, cerov, zhong}.
Moreover, some theoretical predictions relating spectral and
diffusive properties  have been made
\cite{guarneri}. These
predictions can now be tested analytically in the present model,
which will be the subject of a forthcoming paper. 
A second line of future investigation is a perturbation theory
around the exactly solvable point  $g_{\rm loc}=g_{\rm nloc}$. This
introduces a $2\pi-$periodic displaced delta-function with a
weight proportional to $g_{\rm loc}-g_{\rm nloc}$, a quantity which may serve
as small parameter of the perturbative expansion. It is possible that 
localized states appear as soon as $g_{\rm loc}-g_{\rm nloc}$
becomes nonzero, in which case the localized regime would be
perturbatively
accessible.

\begin{acknowledgments}
I wish to thank N. Andrei and K.-H. H\"ock for fruitful discussions
and the Referees for important hints.
\end{acknowledgments}


\begin{thebibliography}{12}


\bibitem{aa} S.~Aubry and G.~Andr\'e, Ann. Israel Phys. Soc. {\bf 3},
  133 (1980)

\bibitem{yuan} H.Q.~Yuan, U.~Grimm, P.~Repetowicz and M.~Schreiber,
Phys. Rev. B {\bf 62}, 15569 (2000)

\bibitem{cerov} V.Z.~Cerovski, M.~Schreiber and U.~Grimm,
Phys. Rev. B {\bf 72}, 054203 (2005)

\bibitem{mandel} S.E.~Mandel and R.~Lifshitz,
Phil. Mag. {\bf 86}, 759 (2006)


\bibitem{exp} Z.M.~Stadnik, D.~Purdie, M.~Garnier, Y.~Baer, A.P.~Tsai,
  A.~Inoue, K.~Edagawa, S.~Takeuchi and K.H.J.~Buschow,
Phys. Rev. B {\bf 55}, 10938 (1997)\\ 
J.~Delahaye, T.~Schaub, C.~Berger and Y.~Calvayrac, 
 Phys. Rev. B {\bf 67}, 214201 (2003)\\
Y.K.~Kuo, K.M.~Sivakumar, H.H.~Lai, C.N.~Ku, S.T.~Lin and A.B.~Kaiser, 
 Phys. Rev. B {\bf 72}, 054202 (2005)

\bibitem{theor}
J.B.~Sokoloff and J.V.~Jose, Phys. Rev. Lett. {\bf 49}, 334 (1982)\\  
C.M.~Soukoulis and E.N.~Economou, Phys. Rev. Lett. {\bf 48}, 1043 (1982)\\ 
D.R.~Grempel, S.~Fishman and R.E.~Prange,
 Phys. Rev. Lett. {\bf 49}, 833 (1982)\\
 M.~Kohmoto, L.P.~Kadanoff and C.~Tang, 
Phys. Rev. Lett. {\bf 50}, 1870 (1983)\\  
S.~Ostlund, R.~Pandit, D.~Rand, H.J.~Schellnhuber and E.D.~Siggia, 
Phys. Rev. Lett. {\bf 50}, 1873 (1983)



\bibitem{loy} P.~Lloyd, J. Phys. C {\bf 2}, 1717 (1969) 

\bibitem{zhong} J.X.~Zhong and R.~Mosseri, 
J. Phys. Cond. Mat. {\bf 7}, 8383 (1995)



\bibitem{guarneri} I.~Guarneri, Europhys. Lett. {\bf 10}, 95 (1989)\\ 
R.~Ketzmerick, G.~Petschel and T.~Geisel,
Phys. Rev. Lett. {\bf 69}, 695 (1992)\\
F.~Pi\'echon, Phys. Rev. Lett. {\bf 76}, 4372 (1996)\\ 
R.~Ketzmerick, K.~Kruse, S.~Kraut and T.~Geisel,
Phys. Rev. Lett. {\bf 79}, 1959 (1997)



\end{thebibliography}
\end{document}